\documentclass[12pt,preprint]{aastex}

\newcommand{\bmax}{\beta_{\rho{\rm max}}}
\newcommand{\rmax}{\rho_{\rm max}}
\newcommand{\bl}[1]{\mbox{\boldmath$ #1 $}}

\slugcomment{accepted by Astrophysical Journal Letters}

\shorttitle{Magnetized Cloud Fragmentation}
\shortauthors{Kudoh \& Basu}

\begin{document}

\title{Three-dimensional Simulation of Magnetized Cloud Fragmentation 
Induced by Nonlinear Flows and Ambipolar Diffusion}

\author{Takahiro Kudoh\altaffilmark{1}
and Shantanu Basu\altaffilmark{2}}

\altaffiltext{1}{Division of Theoretical Astronomy, 
National Astronomical Observatory of Japan, 
Mitaka, Tokyo 181-8588, Japan; kudoh@th.nao.ac.jp.}
\altaffiltext{2}{Department of Physics and Astronomy, 
University of Western Ontario, London, 
Ontario N6A 3K7, Canada; basu@astro.uwo.ca.}

\begin{abstract}
We demonstrate that the formation of collapsing cores 
in subcritical clouds is accelerated by nonlinear flows, by
performing three-dimensional non-ideal MHD simulations. 
An initial random supersonic (and trans-Alfv\'enic) turbulent-like
flow is input into a self-gravitating 
gas layer that is threaded by a uniform magnetic 
field (perpendicular to the layer) such that the initial mass-to-flux 
ratio is subcritical.
Magnetic ambipolar diffusion occurs very rapidly initially due to the
sharp gradients introduced by the turbulent flow. It subsequently occurs
more slowly in the traditional near-quasistatic manner, but in regions
of greater mean density than present in the initial state. 
The overall timescale for runaway growth of the first core(s) is 
several $\times\, 10^6$ yr, even though
previous studies have found a timescale of 
several $\times\, 10^7$ yr when starting with linear perturbations and similar
physical parameters.
Large-scale supersonic flows exist in the cloud and provide an
observationally testable distinguishing characteristic from 
core formation due to linear initial perturbations.
However, the nonlinear flows have decayed sufficiently that
the relative infall motions onto the first core are subsonic,
as in the case of starting from linear initial perturbations.
The ion infall motions are very similar to those of neutrals; 
however, they lag the neutral infall 
in directions perpendicular to the mean magnetic field direction and lead 
the neutral infall in the direction parallel to the mean magnetic field.

\end{abstract}


\section{Introduction}

In the standard model of low-mass star formation, molecular clouds have
an initially subcritical mass-to-flux ratio,
and spend a relatively long time ($\sim 10^7$ years) undergoing quasi-static 
fragmentation by ambipolar diffusion until cores come to be supercritical.
After a core becomes supercritical, it collapses on a dynamical timescale
($\sim 10^6$ years) to form either one or a small multiple system of 
stars \citep[e.g.,][]{shu87,mos91}. Models of nonaxisymmetric
fragmentation by this mechanism have been presented recently in
several papers \citep[][hereafter KBOY]{ind00,bas04,cio06,kud07}.
This scenario is supported by the observed relatively low efficiency
of star formation in molecular clouds, i.e., only $1-5\%$ for 
clouds as a whole but several times greater within cluster-forming cores
\citep[see][]{lad03}.
An alternative model is that star formation begins in clouds with
supercritical mass-to-flux ratio that have additional support due to 
turbulence \citep[e.g.,][]{nak98,har01}.
In this model, star formation occurs relatively rapidly,
as turbulence dissipates over a dynamical timescale,
and the large-scale magnetic field plays a minor role.
The rapid star formation model is supported by the observational
results that the age spreads of young stars in nearby 
molecular clouds are often a few $\times \,10^6$ years \citep{elm00,har01,har03},
and the low fraction of clouds that are observed to be in a pre-star-formation
state.

Recently, \citet{li04} and \citet{nak05} have modeled
mildly subcritical clouds and shown that the timescale of 
cloud fragmentation is reduced by supersonic turbulence. They showed this by
performing two-dimensional simulations in the thin-disk approximation.
Such a model can explain both relatively rapid star formation
and the relatively low star formation efficiency in molecular clouds
that is not well explained if star formation starts from a supercritical cloud.
In this Letter, we study the three-dimensional extension of the
\citet{li04} model, by including a self-consistent calculation 
of the vertical structure and dynamics of the cloud. 
We confirm that 
supersonic flows can significantly shorten the timescale of core formation in 
three-dimensional subcritical clouds, 
and clarify how the scaling of various physical quantities allows this 
to occur.

\section{Numerical Model}

We solve the three-dimensional magnetohydrodynamic 
(MHD) equations including self-gravity and ambipolar diffusion,
assuming that neutrals are much more numerous than ions.
Instead of solving a detailed energy equation, we assume isothermality
for each Lagrangian fluid particle \citep{kud03,kud06}.
For the neutral-ion collision time and associated quantities,
we follow \citet{bas94}. The basic equations 
are summarized in KBOY.

As an initial condition, we assume hydrostatic equilibrium of 
a self-gravitating one-dimensional cloud along the $z$-direction in
a Cartesian coordinate system $(x,y,z)$.
Though nearly isothermal, a molecular cloud is usually surrounded
by warm material, such as neutral hydrogen gas.
Hence, we assume that the initial sound speed makes a transition
from a low value $c_{s0}$ to a high value $c_{sc}$ at
a distance $z=z_c$, with a transition length $z_d$ (see
eq. 16 of KBOY).
We take $c_{sc}^2=10\,c_{s0}^2$, $z_c=2H_0$, and $z_d=0.1H_0$,
where $c_{s0}$ is the initial sound speed at $z=0$,
$H_0=c_{s0}/\sqrt{2\pi G \rho_0}$, and $\rho_0$ is the initial density 
at $z=0$. A numerical solution for the initial density distribution 
shows that it is almost the same as the \citet{spi42}
solution for an equilibrium isothermal layer in the region $0 \leq z \leq z_c$.
We also assume that the initial magnetic field is uniform 
along the $z$-direction.

A set of fundamental units for this problem are
$c_{s0}$, $H_0$, and $\rho_0$. These yield a time 
unit $t_0=H_0/c_{s0}$. The initial magnetic field
strength ($B_0$) introduces a dimensionless free parameter
\begin{equation}
\beta_0 \equiv \frac{8 \pi p_0}{B_0^2} 
= \frac{8 \pi \rho_0 c_{s0}^2}{B_0^2}
= 2\, \frac{c_{s0}^2}{V_{A0}^2} ,
\end{equation}
the ratio of gas to magnetic pressure at $z=0$. In the above relation,
we have also used $V_{A0}\equiv B_0/\sqrt{4 \pi \rho_0}$, 
the initial Alfv\'en speed at $z=0$. 
In the sheet-like equilibrium cloud with a vertical 
magnetic field, $\beta_0$ is related to the mass-to-flux ratio
for Spitzer's self-gravitating layer. The mass-to-flux ratio 
normalized to the critical value is 
$\mu_S \equiv 2\pi G^{1/2} \Sigma_S / B_0$,
where
$\Sigma_S= 2\rho_0 H_0$
is the column density of the Spitzer cloud.
Therefore,
$\beta_0=\mu_S^2$.
Although the initial cloud we used is not exactly 
the same as the Spitzer cloud, $\beta_0$ is a good 
indicator of whether or not the magnetic field 
can prevent gravitational instability.
For the model presented in this Letter, we choose $\beta_0=0.25$ so that 
the cloud is slightly subcritical. For this choice, 
$V_{A0} = 2.8\,c_{s0}$.
Dimensional values of all quantities can be found 
through a choice of $\rho_0$ and $c_{s0}$.
For example, for $c_{s0}=0.2$ km s$^{-1}$ and $n_0=\rho_0/m_n=10^4$ cm$^{-3}$, 
we get $H_0=0.05$ pc, $t_0=2.5 \times 10^5$ yr, and 
$B_0=40\,\mu$G if $\beta_0=0.25$.

The level of magnetic coupling in the partially ionized gas is
characterized by numerical values of the ion number density
$n_{\rm i}$ and neutral-ion collision timescale $\tau_{\rm ni}$.
From eqs. (8) and (9) of KBOY, and using standard values 
of parameters in that paper as well as the values of units used 
above, we find an initial midplane ionization fraction
$x_{\rm i,0} = n_{\rm i,0}/n_0 = 9.5 \times 10^{-8}$ and 
a corresponding neutral-ion collision time $\tau_{\rm ni,0} = 0.11 t_0$. 
The ionization fraction $x_{\rm i}$ and timescale $\tau_{\rm ni}$
at other densities can be found from the initial midplane values given 
that they both scale $\propto \rho^{-1/2}$ \citep{elm79}.

In this equilibrium sheet-like gas layer, we input a nonlinear perturbation
to $v_x$ and $v_y$ at each grid point. Independent realizations of
$v_x$ and $v_y$ are generated
in Fourier space with amplitudes drawn from a Gaussian 
distribution and consistent with power spectrum $v_k^2 \propto k^{-4}$.
Here, $k = (k_x^2+k_y^2)^{1/2}$ is the total wavenumber. 
Our adopted power spectrum means that the large scale has more energy 
than the small scale. 
This perturbation is consistent with that implemented in \citet{li04}.
The rms value of the initial velocity perturbation in physical space, $v_a$, 
is about $3c_{so}$, so that $v_a \simeq V_{A0}$ as well.

The method of solution and boundary conditions are described 
by KBOY \citep[see also][]{kud99,oga04}.
The computational region is $|x|,|y| \leq 8\pi H_0$ and
$0 \leq z \leq 4H_0$, with a number of grid points for 
each direction $(N_x,N_y,N_z)=(64,64,40)$.

\section{Results}

The top panel of Figure 1 shows the time evolution of the maximum density
$\rmax$ at $z=0$.
The simulation is stopped when $\rmax = 30\rho_0$.
The solid line shows the result when we input an initially nonlinear 
supersonic perturbation to the $\beta_0=0.25$ model. 
The dashed line ($\beta_0=0.25$) and the dash-dotted line ($\beta_0=4$) 
show the result in the case of the small (linear) initial perturbation 
that we have applied in KBOY. 
This figure shows that the timescale
of collapsing core formation for the nonlinear perturbation case is much shorter 
than that for the linear perturbation case, when $\beta_0$ is the same.
Even when the initial cloud is subcritical ($\beta_0=0.25$), the core
formation occurs on almost the same timescale as that of 
the supercritical ($\beta_0=4$) linear perturbation case
\footnote{When the initial cloud is supercritical and the perturbation is supersonic, 
the collapsing core formation happens quickly, at $t \simeq t_0$, from the initial flow. 
This may be {\it too} rapid to agree with observations.}.
The bottom panel of Figure 1 shows the time evolution of the maximum value of 
density at $z=0$ ($\rmax$) and $\beta$ at the location of maximum density ($\bmax$).
At first, $\bmax$ increases rapidly 
up to $\sim 0.9$ due to rapid ambipolar diffusion in the highly compressed
regions caused by the initial supersonic perturbation. 
However, there is enough stored magnetic energy in the compressed region
that it rebounds and starts oscillations, with $\bmax$ around $0.7$
and increasing gradually. Eventually, $\bmax$ becomes 
$> 1$ and the dense region collapses to form a core.
This figure implies that $\beta$ is a good indicator 
to see whether a subregion of the cloud is supercritical or not. 
The evolution of $\rmax$
confirms that there is an initial compression followed
by a rebound to a lower density (still greater than the initial background
value) and subsequent oscillations until a runaway collapse starts when
continuing ambipolar diffusion has created a region with $\beta > 1$.

Figure 2 shows an image of the logarithmic density at the last snapshot
($t=20.5t_0$, when $\rmax=30\rho_0$) for the nonlinear perturbation case of 
the subcritical cloud ($\beta_0=0.25$).
The top panel shows the cross section at $z=0$, and the bottom panel 
shows the cross section at $y=-5.9H_0$. The value of $y$ for the bottom
panel is chosen so that the vertical cut passes through the maximum
density point. A collapsing core is located in the 
vicinity of $x=-20H_0,y=-5H_0$.
The size of the core is similar to that created by linear initial 
perturbations (see Fig. 2 in KBOY), although the shape is
notably less circular.

The top panel of Figure 3 shows the density and $x$-velocity
of neutrals and ions
along an $x$-axis cut at $y=-5.9H_0, z=0$, taken from the snapshot
illustrated in Figure 2.
The $x$-velocities show infall motion toward the center of the core,
although the core itself is moving with nonzero negative $x$-velocity.
The relative infall speed to the core is subsonic and about $0.35c_{s0}$.
It is comparable to the case of
initial linear perturbation (see Fig. 9 in KBOY). 
However, there are systematic motions throughout the simulation region
that are still supersonic at this time, even though the initial turbulent
energy has decayed somewhat. This is qualitatively distinct
from the corresponding case with linear perturbations,
and should be observationally testable. 
For the single core that is formed in this simulation, the
systematic core motion is about $0.5c_{s0}$ in the $x$-direction. 
However, it can be even larger in other realizations. 
The ion velocity $\bl{v}_i$ can be related to the neutral velocity
$\bl{v}$ and the magnetic acceleration $\bl{a}_M$ in the assumed limit
of low ion inertia by the relation
\begin{equation}
\bl{v}_i = \bl{v} + \bl{a}_M\, \tau_{\rm ni}.
\end{equation}
Figure 3 reveals that the ion infall motions are smaller in magnitude than the 
neutral motions due to the retarding magnetic acceleration; however
the relative drift in the $x$-direction between ions and neutrals 
is typically within $0.05c_{s0}$.  
In order to see the force balance in the core, we plot
the $x$-accelerations in the bottom panel of Figure 3 
along the same $x$-axis as the top panel.
These values are normalized by the computational units.
The inward gravitational acceleration ($a_G$) dominates around the core.
The acceleration from the magnetic force ($a_M$) and 
thermal pressure force ($a_T$) resist the contraction.
The net acceleration ($a_{\rm net}$) is working toward 
the center of the core.

The top panel of Figure 4 shows the density and $z$-velocities
along a $z$-axis cut at $x=-20H_0,y=-5.9H_0$ from the snapshot illustrated
in Figure 2. The $z$-velocities also show infall motion toward 
the center of the core. The relative neutral infall speed to the core 
is also subsonic and about $-0.35c_{s0}$. The ion velocity in the vertical
direction is almost identical to the neutral velocity but actually 
very slightly {\it greater} in magnitude at the peak.
The bottom panel of Figure 4 shows the $z$-accelerations along 
the same $z$-axis as in the top panel. In the $z$-direction, 
$a_G$ is nearly balanced by $a_T$, and $a_M$ is almost negligible,
{\it but working in the same direction as gravity}.
The net acceleration ($a_{\rm net}$) is also working toward 
the center of the core.
The inward pointing vertical magnetic acceleration associated 
with an hourglass-type
magnetic field morphology explains why the ion velocity in the 
$z$-direction is greater in magnitude than the neutral velocity.

\section{Discussion and Conclusions}

Our three-dimensional MHD simulations have shown that the supersonic nonlinear 
flows significantly reduce the timescale of collapsing core formation 
in subcritical clouds. It is of order several $\times \,10^6$ years 
for typical parameters, or $\sim 10$ times less than 
found in the linear initial 
perturbation studies of \cite{bas04}, \cite{cio06}, and KBOY. 
Our supersonic perturbations are also trans-Alfv\'enic, in agreement with
analysis of observed magnetic field strengths \citep{bas00}. 
It is also in line with
the theoretical result of \citet{kud03,kud06} that turbulent 
clouds will settle into
a trans-Alfv\'enic state when global motion/expansion is allowed, 
even if driven with turbulence that is initially super-Alfv\'enic.

To see how accelerated ambipolar diffusion can occur, we note
that the magnetic induction equation (see, e.g., eq. 3 of KBOY)
and our assumptions of ionization balance 
can be used to estimate the diffusion time $\tau_d \propto
\rho^{3/2}L^2/B^2$, where $L$ is the gradient length scale introduced
by the initial turbulent compression and $B$ is the magnetic field
strength. 
Because the compression by the nonlinear flow is nearly one-dimensional, 
the magnetic field scales roughly as $B \propto L^{-1}$ within the
flux freezing approximation. 
If the compression is rapid enough that vertical hydrostatic
equilibrium cannot be established (unlike in previous calculations
using the thin-disk approximation), then $\rho \propto L^{-1}$ as well
(i.e., one-dimensional contraction without vertical settling), 
and $\tau_d \propto L^{5/2}$. This means that diffusion can occur 
quickly (and lead to a rapidly rising value of $\beta$) 
if the turbulent compression creates small values of $L$.
If diffusion is so effective during the first turbulent compression
that a dense region becomes magnetically supercritical, then it
will evolve directly into collapse. Alternately, the stored magnetic
energy of the compressed (and still subcritical) region may lead to a 
reexpansion of the 
dense region. The timescale for this, in the flux-freezing limit,
is the Alfv\'en time $\tau_A \propto L\rho^{1/2}/B$, which scales
$\propto L^{3/2}$ for the above conditions. Thus, $\tau_d$ decreases
more rapidly than $\tau_A$, and sufficiently small turbulent-generated
values of $L$ may lead to enough magnetic diffusion that collapse occurs 
before any reexpansion can occur.
See \citet{elm07} for some similar
discussion along these lines. 
Ultimately, whether or not reexpansion
of the first compression can occur depends on the strength of the
turbulent compression, mass-to-flux ratio of the initial cloud,
and neutral-ion collision time.
In a preliminary study of such effects, we found that keeping 
$v_a$ and $\beta_0$ fixed but allowing significantly poorer 
or stronger neutral-ion coupling yielded 
differing results. For $\tau_{\rm ni,0} = 0.3 t_0$ (poorer coupling), 
a collapsing
core formed due to the first compression, at $t\simeq 1.4t_0$,
while for $\tau_{\rm ni,0} = 0.05 t_0$ it happened after more 
oscillations than in our standard model, at $t\simeq 85t_0$.
A full parameter study will be required to elucidate.

If reexpansion of the initial compression does occur, as in the standard model
presented in this paper, then there is enough time for the vertical
structure to settle back to near-hydrostatic equilibrium, in which 
case $B \propto \rho^{1/2}$. Since the compressed and reexpanded region
executes oscillations about a new mean density, it is convenient to 
analyze the scalings in terms of the density $\rho$. The diffusion time
now scales as $\tau_d \propto \rho^{-1/2}$.
This yields a scaling of $\tau_d$
that is the traditionally used one (and is satisfied by 
design in the thin-disk approximation). However, the diffusion
occurs more rapidly than it would in the initial state due to the elevated
value of $\rho$ in the compressed but oscillating region (see
Figure 1). 

\acknowledgments

SB was supported by a grant from NSERC.
Numerical computations were done mainly on the VPP5000
at the National Astronomical Observatory.

\clearpage

\begin{figure}
\epsscale{.80}
\plotone{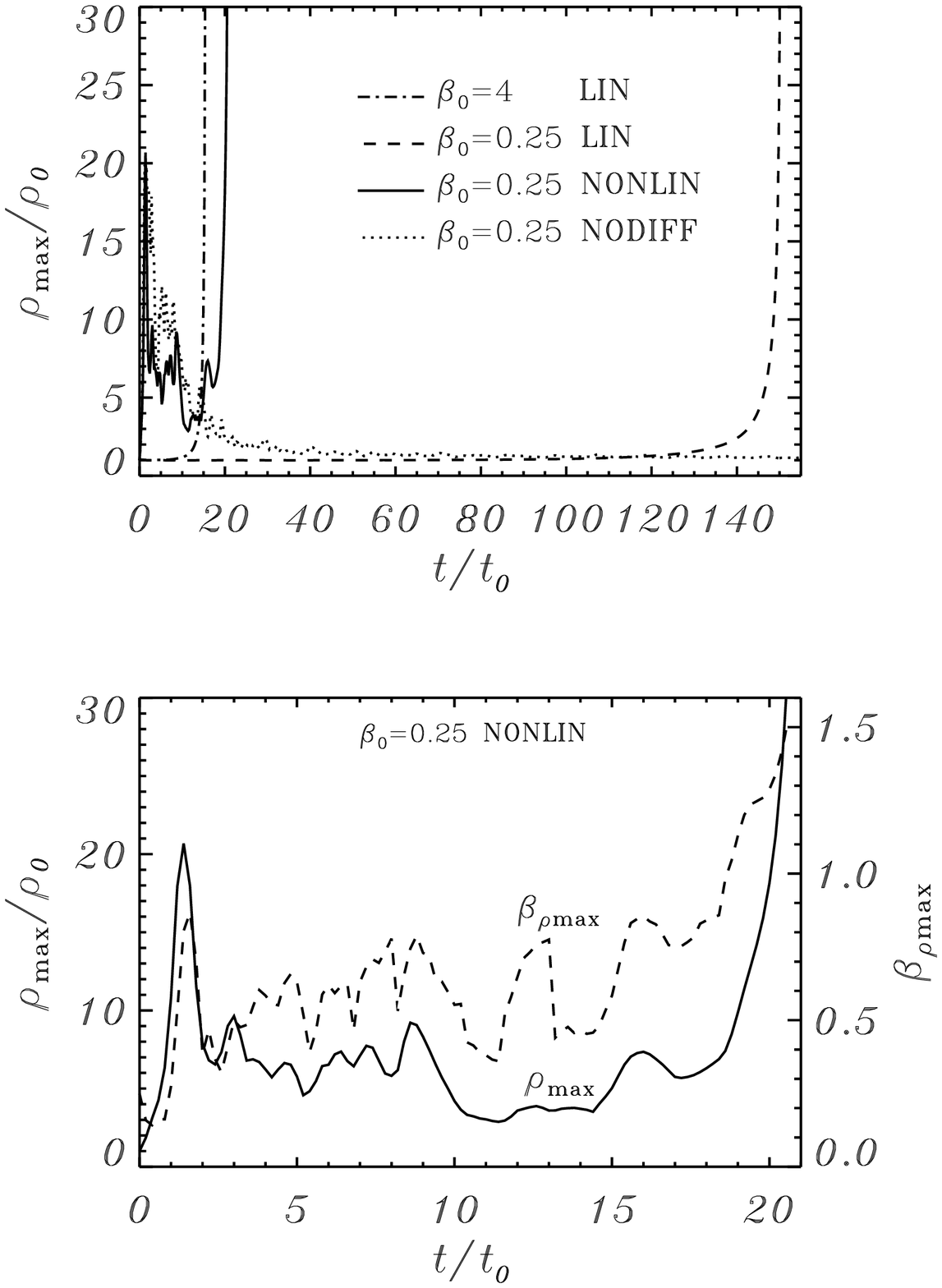}
\caption{
{\it Top}: the time evolution of maximum densities at $z=0$.
The solid line shows the evolution for an initially nonlinear 
supersonic perturbation and $\beta_0=0.25$. The dashed line ($\beta_0=0.25$) and 
the dash-dotted line ($\beta_0=4$) show the evolution for models with
a linear initial perturbation as calculated by KBOY.
The dotted line shows the evolution 
for an initially nonlinear supersonic perturbation and $\beta_0=0.25$,
but without ambipolar diffusion.
{\it Bottom}: evolution of the maximum density (solid line) at $z=0$ of 
the simulation box for the model with $\beta_0=0.25$ and 
nonlinear initial perturbation, and evolution of $\beta$ at 
the location of maximum density (dashed line).
}
\end{figure}

\clearpage

\begin{figure}
\plotone{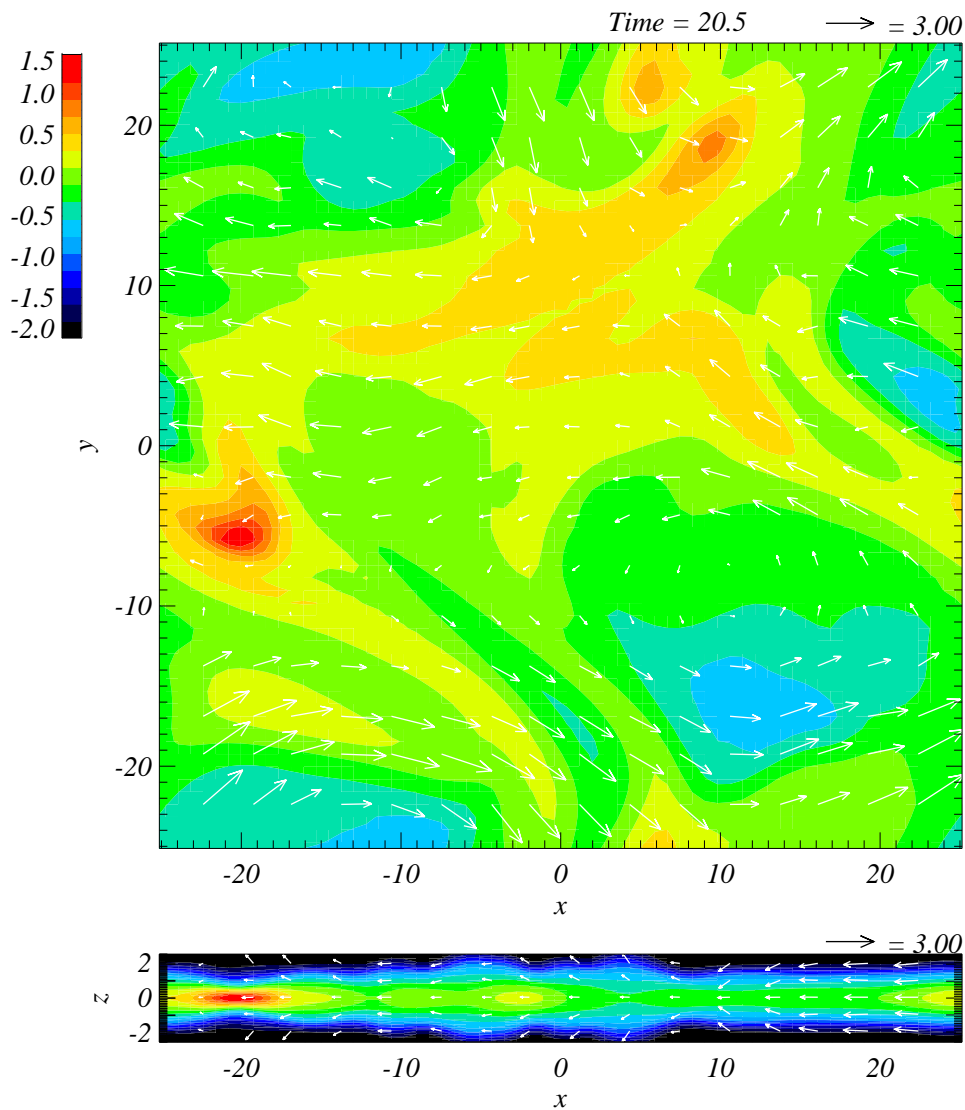}
\caption{
Logarithmic density image at $t=20.5t_0$ for the nonlinear 
perturbation case of the subcritical cloud ($\beta_0=0.25$).
The top panel shows the cross section at $z=0$, and the bottom panel 
shows the $x-z$ cross section at $y=-5.9H_0$.
An mpeg animation of the evolution of the density image from $t=0$ to $t=20.5t_0$ is
available online.
}
\end{figure}

\begin{figure}
\plotone{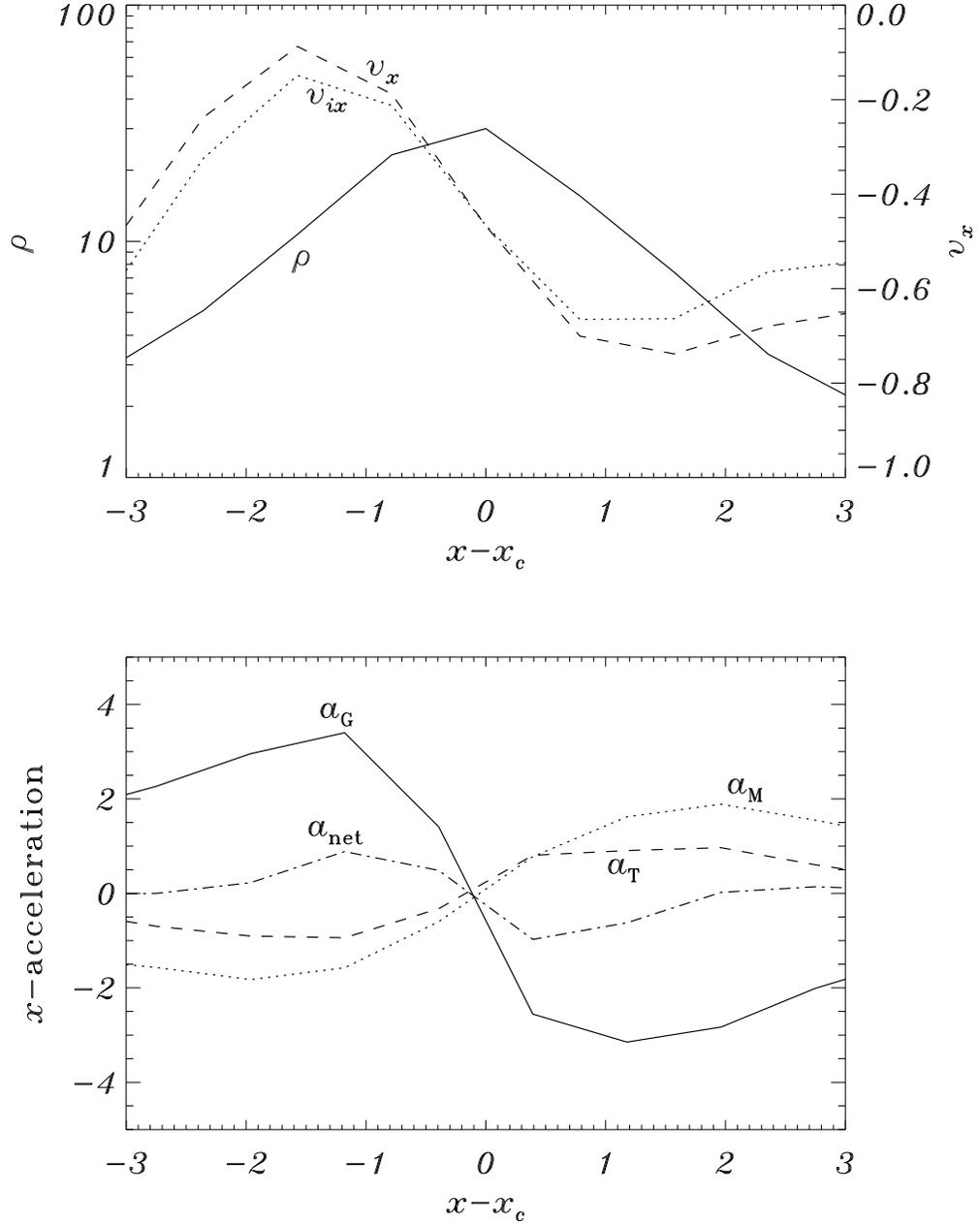}
\caption{
{\it Top}: the density (solid line), $x$-velocity of neutrals
(dashed line), and $x$-velocity of ions (dotted line) along an $x$-axis cut 
at $y=-5.9H_0,z=0$, in the snapshot shown in Fig. 2. 
The $x$-positions are measured by offset from
$x_c=-20H_0$, which is the maximum density point for the core.
{\it Bottom}: the $x$-accelerations along 
the same $x$-axis as the top panel.
The solid line is the gravitational acceleration ($a_G$),
the dotted line is the magnetic acceleration ($a_M$),
the dashed line is acceleration due to thermal pressure ($a_T$),
and the dash-dotted line is the net acceleration ($a_{\rm net}$). 
All quantities are normalized by the computational units.
}
\end{figure}

\begin{figure}
\plotone{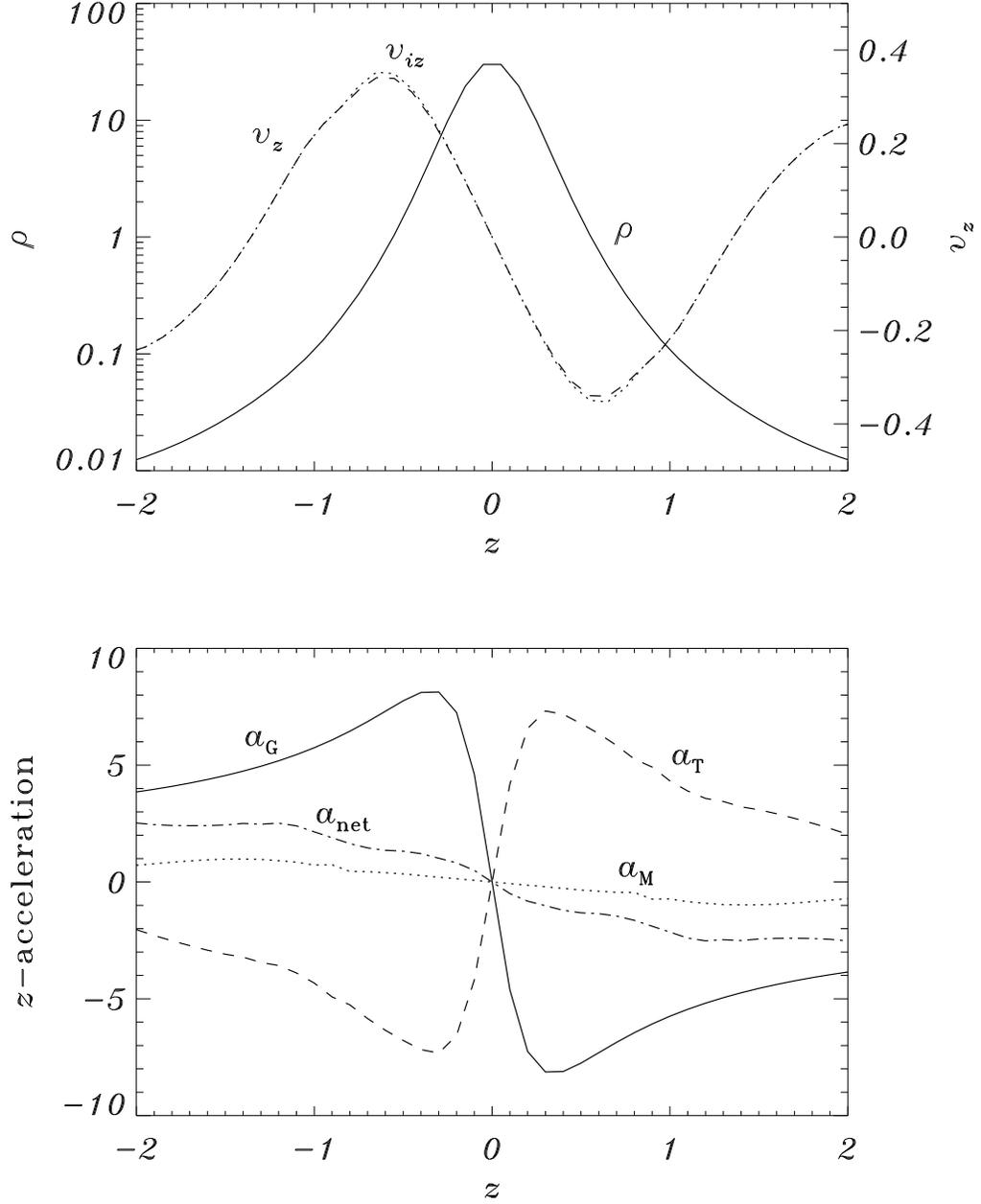}
\caption{
{\it Top}: the density and $z$-velocity of neutrals and ions 
along a $z$-axis cut at $x=-20H_0,y=-5.9H_0$,
in the snapshot shown in Fig. 2.
{\it Bottom}: the $z$-accelerations along 
the same $z$-axis as the top panel. 
The line styles are the same as those in Fig. 3.
}
\end{figure}

\end{document}